\documentclass[12pt]{iopart}

\usepackage{iopams}
\usepackage{amsmath2}
\usepackage{graphicx}

\begin{document}

\title{Lorentz-violation-induced arrival delays of cosmological particles}
\author{Uri Jacob and Tsvi Piran}
\address{Racah Institute of Physics, The Hebrew University, Jerusalem, Israel}
\eads{\mailto{uriyada@phys.huji.ac.il}, \mailto{tsvi@phys.huji.ac.il}}

\begin{abstract}
We point out that previous studies of possible Lorentz-violating effects in astronomical time-of-flight data did not take into account the entire implications of the universe's cosmological expansion. We present the derivation of the accurate formulation of the problem and show that the resulting correction of the limits on Lorentz violation is significant.
\end{abstract}

Lorentz invariance violation (LIV) arises in various frameworks and theories of quantum-gravity \cite{QG1,QG2}. In the context of quantum-gravity, spacetime may have a non-trivial small-scale structure, so the laws of physics may be altered around the Planck scale ($E_{pl}=\sqrt{\hbar c^5/G}\sim1.2\times10^{28}eV$). At lower energy scales small deviations from Lorentz invariance may be felt. As the effects of LIV may become substantial for high-energy particles, there has been a growing interest in testing LIV by means of time-of-flight measurements of astronomical particles. Time-of-flight analyses provide the most generic method of testing LIV. Astronomical sources provide, in addition to high observable energies, large propagation distances that amplify the small LIV effects. The use of cosmological sources to gain insight on possible modifications to particles' flight times due to LIV requires a careful consideration of the universe's cosmological expansion. \par

Within the LIV phenomenology massless particles may have energy-dependent speeds, and so high-energy particles may arrive with a delay (or an early arrival, but we use the term delay here to imply either cases) compared to low-energy particles emitted at the same instant. The first attempt to quantify limits on LIV by analyzing energy-dependent features in the light curves of gamma-ray bursts (GRBs) \cite{Schaefer} used an approximation neglecting the cosmological expansion. Later, Ellis et al. \cite{Ellis,Ellis2} performed a systematic LIV study based on a combined analysis of an ensemble of GRBs at cosmological distances up to $z>6$. They were able to place a robust lower bound on the energy scale of linear LIV at $\sim10^{-3}E_{pl}$. Ellis et al. introduced \cite{Ellis} a formulation that included the effect of the cosmological expansion, and several groups employed their formulae to perform additional time-of-flight analyses of GRBs \cite{Boggs,Cheng,Gogberashvili,Lamon,Scargle}. The different studies produced quantitative limits on the energy scale of LIV and also predicted the scales that will be examinable with future experiments. In this short note we point out a small but important mistake in the calculation of the cosmological LIV delays, presented in \cite{Ellis} and used in the subsequent studies. We present the derivation of the LIV delay expression we introduced in \cite{ourNature} (see also \cite{Biesiada,Bolmont}) and show that this expression entails a significant correction, which changes by a factor of order $(1+z)$ the LIV limits derived earlier. \par

We consider a general model in which there is a break-down of Lorentz symmetry at some very high energy scale (possibly the quantum-gravity energy scale), which we denote as $\xi E_{pl}$. When examining particles with energies much smaller than the symmetry breaking scale, we may regard only the leading order correction. If the leading LIV correction is of order $n$, the dispersion relation of massless particles can be generically approximated by:
\begin{equation}
E^2-p^2c^2\simeq\pm p^2c^2\left(\frac{pc}{\xi E_{pl}}\right)^n,
\end{equation}
where $pc$ may be freely substituted with $E$ in the LIV term, as we are only interested in the leading order correction. $c$, the 'conventional' speed of light constant, is, in this framework, the speed of low-energy photons. We choose for clarity to use the '$-$' case of subluminal motion for the rest of our derivation. \par

Two particles, that are emitted simultaneously from a source and have different propagation speeds, will arrive on Earth at different times. If the source is at a cosmological distance, then as a result of the universe's expansion, the proper (physical) distances traveled by the particles will also differ. A length that is by definition always fixed between the source and the observer (provided they move together with the universe's expansion) is the 'comoving distance'. In order to determine the cosmological LIV delay between the two particles we have to inspect their comoving trajectories. The comoving trajectory of a particle is obtained \cite{Maria} by writing its Hamiltonian in terms of the comoving momentum:
\begin{equation}
\mathcal{H}=\frac{pc}{a}\sqrt{1-\left(\frac{pc}{a\xi E_{pl}}\right)^n},
\end{equation}
where $a=\frac{1}{1+z}$ is the cosmological expansion factor. Assuming that the standard relation $v=d\mathcal{H}/dp$ holds and neglecting higher order LIV corrections, we write the comoving path as:
\begin{equation}
x(t,p)=\int_0^t\frac{c}{a(t')}\left(1-\frac{1+n}{2}\left(\frac{pc}{a(t')\xi E_{pl}}\right)^n\right)dt',
\end{equation}
where $p$ here is a constant, equal to the present-day momentum. Rewriting this, we obtain the comoving distance traversed by a massless particle, emitted at redshift $z$ and traveling up to redshift $0$:
\begin{equation}\label{LIDtrajectory}
x(z,E_0)=\frac{c}{H_0}\int_0^z\left(1-\frac{1+n}{2}\left(\frac{E_0}{\xi E_{pl}}\right)^n(1+z')^n\right)\frac{dz'}{\sqrt{\Omega_m(1+z')^3+\Omega_\Lambda}}.
\end{equation}
$E_0$ is the redshifted particle energy measured at present. $\Omega_m$, $\Omega_\Lambda$ and $H_0$ are the cosmological parameters evaluated today. We examine a low-energy photon, that was emitted at redshift $z$ and reaches us at redshift $0$, and a highly energetic one, that was also emitted at redshift $z$ and arrives with a delay at redshift $-{\scriptstyle\Delta}z$. The comoving distance, traveled by both particles, emitted from the same source and reaching Earth, is the same. Equating the two paths and taking again only the leading order LIV corrections yields:
\begin{equation}
\int_{-{\scriptscriptstyle\Delta}z}^0\frac{dz'}{\sqrt{\Omega_m(1+z')^3+\Omega_\Lambda}} =\int_0^z\left(\frac{1+n}{2}\left(\frac{E_0}{\xi E_{pl}}\right)^n(1+z')^n\right)\frac{dz'}{\sqrt{\Omega_m(1+z')^3+\Omega_\Lambda}}.
\end{equation}
Since for all delays that may be considered ${\scriptstyle\Delta}z$ is a very small number, we neglect second order corrections in ${\scriptstyle\Delta}z$ and arrive at the expression for the LIV time delay of a cosmological high-energy massless particle:
\begin{equation}\label{LIDdelay}
{\scriptstyle\Delta}t=\frac{{\scriptstyle\Delta}z}{H_0}=\frac{1+n}{2H_0}\left(\frac{E_0}{\xi E_{pl}}\right)^n\int_0^z\frac{(1+z')^n dz'}{\sqrt{\Omega_m(1+z')^3+\Omega_\Lambda}}.
\end{equation}
It is worth noting that the same expression is valid for an ultrarelativistic massive particle, such as an energetic neutrino \cite{ourNature}, if the correction term in the dispersion relation due to the particle's mass is smaller than the correction due to LIV. \par

In \cite{Ellis} Ellis et al. derived the delay between two photons arriving from redshift $z$ with present-day energy difference ${\scriptstyle\Delta}E$ for the case of a linear LIV induced correction to the dispersion relation. Their result in our notations, ${\scriptstyle\Delta}t=H_0^{-1}\frac{{\scriptstyle\Delta}E}{\xi E_{pl}}\int_0^z\frac{dz'}{\sqrt{\Omega_m(1+z')^3+\Omega_\Lambda}}$, lacks a factor of $(1+z)$ in comparison with our expression (for $n=1$). The reason for this is that the derivation in \cite{Ellis} equates the proper distances that the two photons traverse by the time they reach Earth. As mentioned, this is incorrect - the source and the Earth have a fixed comoving distance between them, but the proper distance varies as the universe expands. While the energetic photon is delayed throughout the path, the universe's expansion factor is very slightly modified, but this slight modification affects the large cosmological distance that the particle must pass before reaching the observer. Hence, this effect accumulates to give a significant difference in the proper distances traveled. A detailed calculation shows that the time delay due to the additional proper distance that the slower particle passes is of the same order of magnitude as the delay produced between the particles by the time they complete the same proper distance. Thus, when inspecting the delays of arrival on Earth of particles propagating from distant cosmological sources, the expansion of the universe during the delay periods cannot be neglected. Another simple illustration of the problem with the derivation in \cite{Ellis} is that the delays produced between the particles at different intervals along the path are merely summed up, without considering the fact that a delay produced further back in the path amounts to a larger delay on Earth. This effect of relativistic dilation introduces a factor of $(1+z)$ into the above integral. \par

Although we are dealing with tiny modifications to conventional physics that produce short delays, the detailed cosmological analysis is necessary and the correction we presented here is significant. For example, using standard cosmology ($H_0=70\frac{km}{sec}Mpc^{-1}$, $\Omega_m=0.3$, $\Omega_\Lambda=0.7$) and referring to the scenario of $n=1$ LIV with $\xi=1$ (the symmetry breaking scale is the Planck energy), we obtain with (\ref{LIDdelay}) for a $1MeV$ photon emitted at $z=1$ a delay of  $4.1\times10^{-5}sec$, in comparison to a delay of $2.8\times10^{-5}sec$ according to the calculation in \cite{Ellis}. For more distant sources the factor between the correct calculation and the erroneous one will be even larger. We find that the required correction of the cosmological LIV delay calculations would serve to extract stronger LIV bounds from the previously analyzed GRB data. For example, given the range of redshifts in the sample of sources used by Ellis et al., this correction implies that the limits are stronger by a factor of $\sim2$.

\end{document}